# Mapping EU fishing activities using ship tracking data

Michele Vespe, Maurizio Gibin, Alfredo Alessandrini, Fabrizio Natale, Fabio Mazzarella, Giacomo C. Osio
European Commission, Joint Research Centre, Italy

*Abstract*: Information and understanding of fishing activities at sea are fundamental components of marine knowledge and maritime situational awareness. Such information is important to fisheries science, public authorities and policy makers. In this paper we introduce a first map at European scale of EU fishing activities extracted using Automatic Identification System (AIS) ship tracking data. The resulting map is a density of points that identify fishing activities. A measure of the reliability of such information is also presented as a map of coverage reception capabilities.

## Introduction

Mapping fishing activities at sea has been investigated using data from different ship tracking systems. Amongst them, terrestrial networks of Automatic Identification System (AIS) receivers offer the possibility to track vessels independently of their flag, at wide scale. Conceived primarily as a collision avoidance and vessel monitoring tool, AIS is nowadays a cornerstone of safety of navigation. In addition, AIS is becoming progressively used to improve Maritime Situational Awareness to detect ship's behaviour anomalies and project the current situation into the future (Pallotta et al., 2013). Historical AIS data have been proved effective in mapping merchant routes (Fernandez et al., 2014) and other activities such as off-shore, research and exploration etc. (Vespe et al., 2015). In Europe (EU Dir 2011/15/EU), even fishing vessels down to 15 m in length are required to be fitted with AIS, and this has a significant impact in understanding the spatial distribution of fishing efforts (e.g., Natale et al., 2015, and Mazzarella et al., 2014) and the economics of the EU fisheries sector (Natale et al., 2016). The knowledge of fishing activities is not only fundamental for fisheries science, but it is also a key element for policy makers, e.g. in planning activities at sea (Maritime Spatial Planning) or in assessing the impact of introducing new Marine Protected Areas.

Following the findings reported in (Natale et al., 2015), the paper introduces for the first time an aggregated and anonymised map of EU fishing activities at European scale. The map is produced by processing AIS historical data in the period September 2014 to September 2015. A spatial coverage of the data is also computed after aggregating the data from multiple providers in order to give an estimate of the reliability of the results in the different areas.

## Data

The data used in this analysis originated from AIS terrestrial networks of receivers and contain information on the time, position, direction and speed of individual vessels above 15 meters length. AIS data format and structure are very similar to the output of a commercially available GPS device. Having a rather high time and space granularity, AIS data require additional care on the shortening of the computational time needed for the implementation of the analysis work flow. The AIS dataset used in our analysis contains fishing vessels only. In order to identify fishing vessels the available vessel identifiers transmitted with AIS (Maritime Mobile Service Identity - MMSI) are linked to the EU fishing fleet register[1] through call sign and name. The output of this process is a set of EU fishing vessels whose dynamic information is later processed. In addition, also information on the primary and secondary gears from the fleet register are stored and subsequently used to isolate specific fishing categories such as trawlers, purse seiners, etc. (ISSCFG, 1980). In this study we analyse fishing activities related to trawlers only (bottom otter, beam and midwater trawls[2]) which represent the largest portion of the EU fishing vessels above 15 meters of length. The choice of this group of actively towed fishing gears was based on the importance of these gears and on the reliability of identifying when a boat is actively fishing rather than steaming (Mazzarella et al., 2014). Once linked to the fleet register, the dataset was anonymised.

---

[1] European Union Fleet Register data (http://ec.europa.eu/fisheries/fleet/)

## Methods

The methodology to map EU fishing activities using AIS data can be summarised in the following steps:

*__Data cleaning__*: the position and speed data relevant to the above vessels are selected and cleaned in case of errors. The data are also decimated to an interval of 5 minutes between consecutive observations. This brings to a total of more than a 150 million messages analysed in this study and 60 million of such messages classified as fishing by the model.

*__Fishing Behaviour Identification__*: the points corresponding to fishing behaviour (see Algorithm 1) are extracted based on the assumption that there is a separation of fishing activities with a steaming speed that is relatively high with respect to fishing speed (Mazzarella et al., 2014).

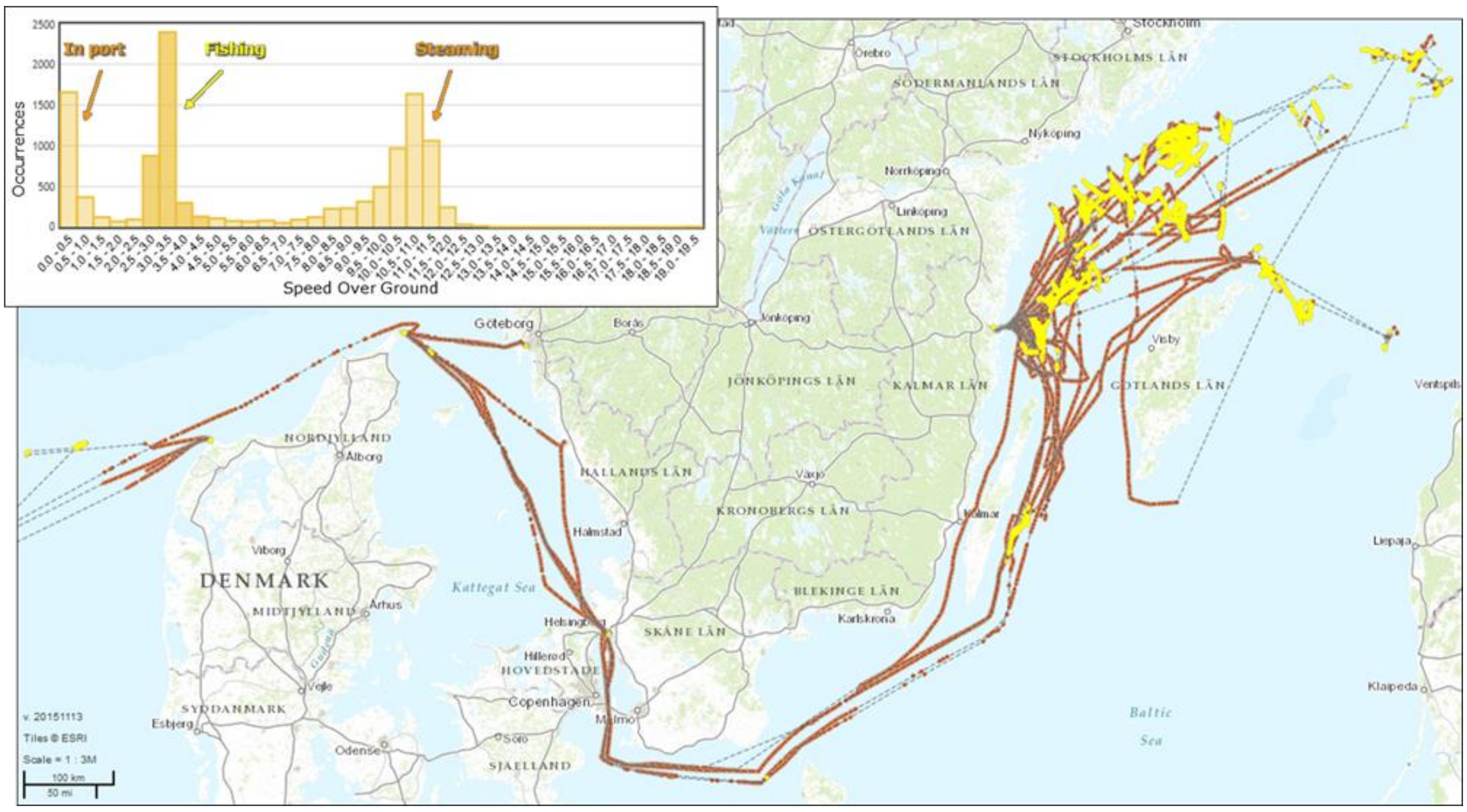

*Figure 1: track and speed profile of a trawler showing three clusters of velocities corresponding to in port, fishing (position highlighted in yellow) and steaming behaviours*

The resulting speed profiles, after excluding the zero-velocity points that relate to messages sent when the vessel is likely to be in a port, show a bi-modal distribution. This is shown in Figure 1, where the track of a trawler is plotted on a map and colour-coded by these speed intervals. The transit legs contain the high speed points, whereas the low speed points around 3.5 knots cluster in the middle of the sea at the far ranges of the track (yellow), and are therefore likely indicating fishing grounds. By applying this approach for all fishing vessels in a specific area, it is possible to understand the fishing grounds and map high intensity fishing areas.

[2] ISSCFG codes Bottom otter trawl (OTB), Beam trawl (TBB), Bottom pair trawl (PTB), Midwater otter trawl (OTM), Midwater pair trawl (PTM), and Multi-rig otter trawl (OTT) (ISSCFG, 1980).

Depending on vessel size, area, fishing gear and many other factors, fishing vessels exhibit specific mean and standard deviation values of the speed bi-modal distributions. For this reason, the identification of fishing behaviour has to be implemented for each individual vessel (Natale et al., 2015). Using a Gaussian Mixture Model (GMM), an unsupervised classification method, we isolated the two main activities distributions and obtained the component parameters. Fishing speed confidence intervals were built for each vessel using the first component mean and standard deviation.

**Algorithm 1: Fishing Behaviour Identification**

```
Require: AIS_messages // AIS messages identified by MMSI and include lat, lon, speed etc.
Require: MMSI_list // List of MMSI# that identify EU fishing vessels
Require: Fishing_points = []

for all MMSI_i ∈ MMSI_list do
// Construct the speed profile for the fishing vessels identified by the MMSI_i
    speed_data_i ← AIS_messages(find(AIS_messages.MMSI = MMSI_i)). speed
// Extract the relevant track for the same vessel
    lat_i ← AIS_messages(find(AIS_messages.MMSI = MMSI_i)).lat
    lon_i ← AIS_messages(find(AIS_messages.MMSI = MMSI_i)).lon
// Find the parameters of the mixture of two Gaussian distributions through EM algorithm
// over speed values greater than 0.5 (this excludes values relevant to vessels in port or close to it)
    [μ_i_1, μ_i_2, ϑ_i_1, ϑ_i_2] ← Expectation_Maximisation (speed_data_i > 0.5)
// Evaluate the fishing speed thresholds for each vessel as ± k standard deviations from the first mode
    [v_th_l, v_th_h] ← μ_i_1 ± k·ϑ_i_1
// Isolate fishing points between ± k standard deviations from the first mode
    Fishing_indexes_i ← find(v_th_l < speed_data_i < v_th_h ))
    Fishing_points ← add [lat_i(Fishing_indexes_i), lon_i(Fishing_indexes_i)]
end for
return Fishing_points
```

_Aggregate results into density maps_: the resulting points classified as fishing are aggregated into 1 $km^2$ cells. It is worth noting that the AIS time decimation allows for the computation of a time-coherent map, where every point corresponds to a 5 minutes fishing activity. In Figure 2, the density of all messages classified as 'fishing' is reported. The map highlights the high intensity fishing areas in EU waters over one year.

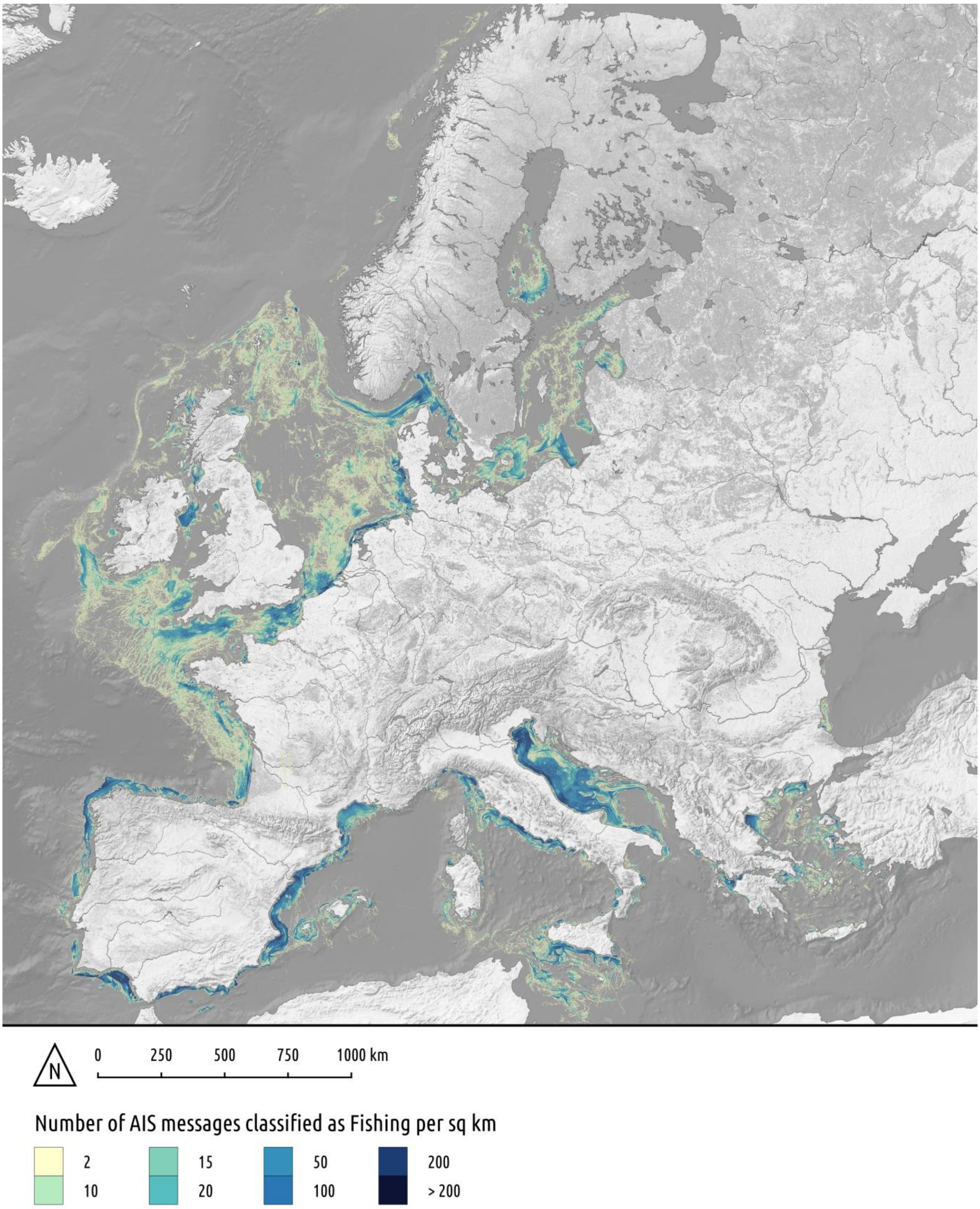

*Figure 2: Density of AIS messages classified as "fishing"*

**Accuracy and validation**

The aggregated fishing intensity value for a specific cell is subject to the completeness of information that can be collected in that area. In our case, since the reception of AIS is related to the radio propagation of the messages transmitted by the vessel, such completeness is mainly linked to the distance to the closest AIS receiving station. Nevertheless, the propagation of AIS messages is also influenced by the atmospheric conditions in the area and significantly varies accordingly.

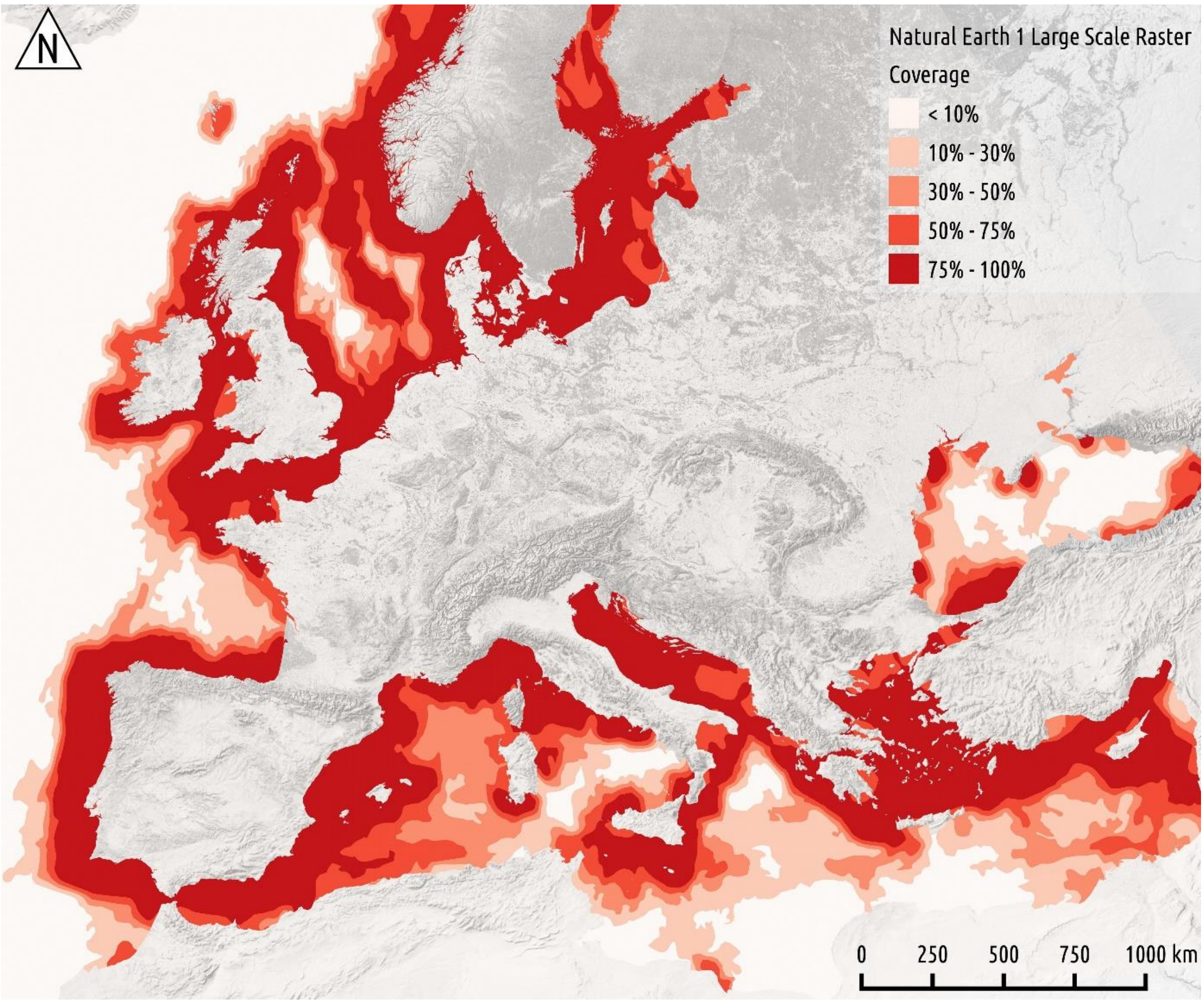

*Figure 3: Spatial coverage map representing the reliability of the results in Figure 2*

The AIS reception spatial coverage in this study is averaged over the AIS dataset period and is shown in Figure 3. This has been calculated for each cell as the ratio between received and expected tracks of vessels traveling at normal cruising speed. The majority of them are merchant vessels that, differently from fishing vessel, follow specific routes. Consequently the trajectories can be derived even with incomplete tracks. A coverage map tells where the information is expected to be complete, and therefore reliable. As an example, low fishing intensity values in areas with low reception capability may be underestimated. Thus, the map of fishing intensity areas has to be used together with the relevant coverage map.

The coverage map shows high coverage in all European coastal waters and over the continental shelf (sea bottom up to 200 m depth). This are the areas of operation of all bottom and – partially - mid water trawls, thus the AIS coverage will be reliable for these gears. In particular, in Figure 4 the results of the density of AIS messages classified as “fishing” is shown over part of the continental shelf. It can be seen that high density areas are located in correspondence of isobaths and follow the bathymetric profile of the area. This is a known behaviour of trawlers that often operate at constant depths. Moreover, it can be observed that the known shipping routes in the areas are not visible, demonstrating that the fishing activity is consistently isolated from the steaming behaviour.

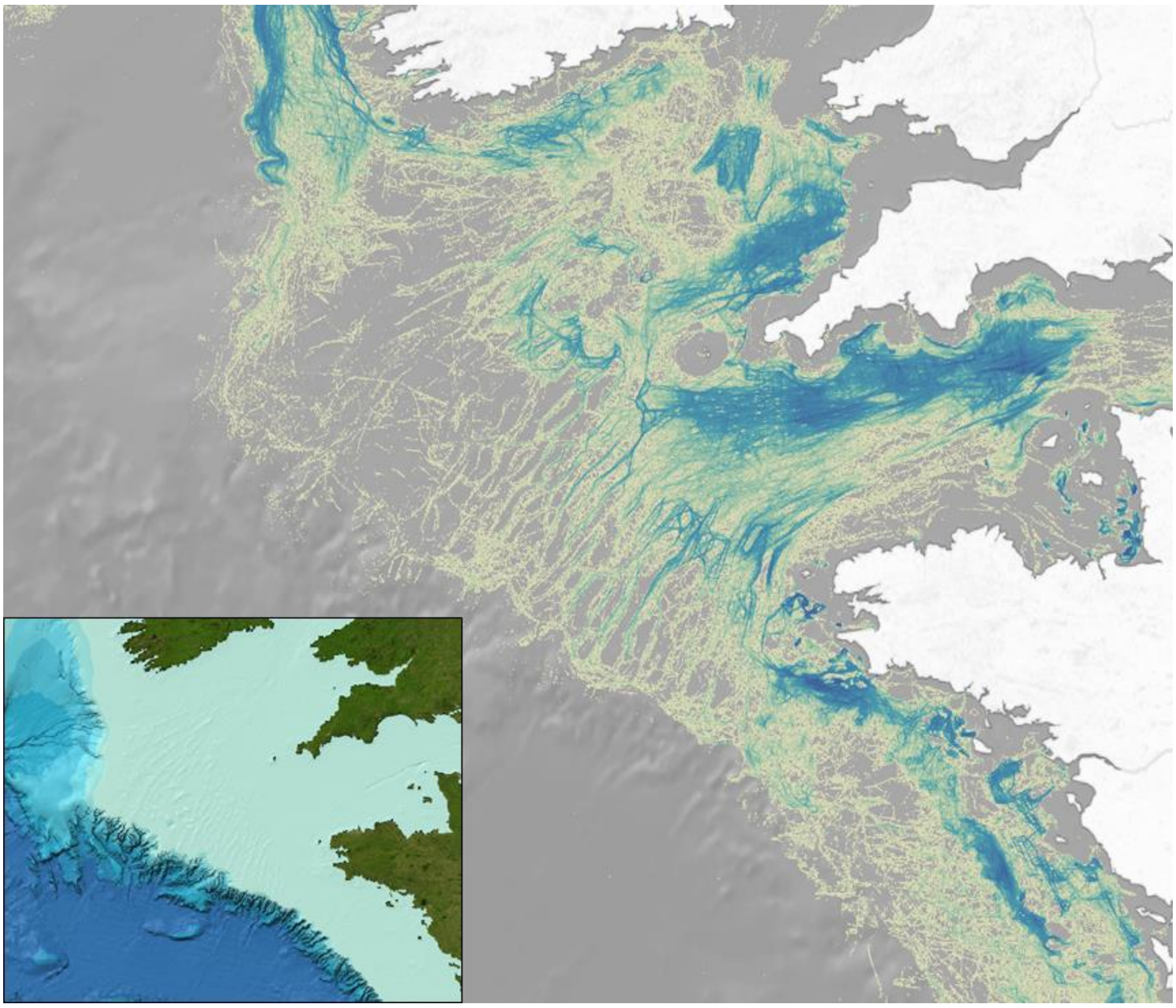

*Figure 4: AIS derived fishing activities over the continental shelf extracted from Figure 2 highlighting the correlation with the bathymetry over the area (bottom-left, from the GEBCO world map 2014, www.gebco.net)*

The coverage map provides reliable information on the accuracy and the precision of the density map, while validation of the machine learning method used in the analysis has to be assessed indirectly. The density map produced represents to our knowledge the first attempt in creating a European level fine scale dataset of fishing intensity. Information about the effort[3] of European fishing vessel is available through the European Data Collection Framework. This information is mostly collected from logbooks and is aggregated by ICES statistical rectangles. ICES statistical rectangles are a bespoke gridded geography introduced by the International Council for the Exploration of the Sea (ICES) in the 1970s and have a size of 1 degree longitude and 0.5 degree latitude.

At the moment, there is not requirement for Member States to submit more detailed fishing activity data, however, it has become increasingly essential to access finer scale fishing information to assess fisheries management measures.

Vessel positioning data is also available in the EU through the Vessel Monitoring Systems (VMS). VMS data is being used to derive high resolution maps of fishing activity on a regional scale by the ICES working group on Spatial Data Fisheries (WGSFD). The WGSFD's main duties are to advice on collating, aggregating, analyse and present spatial fisheries information. The final output of the WGSFD is an annual report (ICES 2015) and data products available to the public. Both the report and the data products produced for the this year contains 2013 data about fishing intensity not calculated and measured as the one produced by the machine learning method presented in this paper.

---

[3] Effort is a more sophisticated measure of fishing intensity and it is constructed as the product of the engine power of the fishing vessel by the amount of time spent at sea fishing and it is measured in *Kilowatt per days at sea*. Fishing intensity is expressed as number of points classified as fishing per square kilometre.

**Conclusions**

The paper has introduced a first map at European scale of EU fishing activities using AIS ship tracking data over one year. For the first time it is possible to compare the intensity of fishing activity, in terms of density of AIS signals, of towed gear at EU level on similar scales.

The first emerging pattern is that all the continental shelf area in the EU Mediterranean countries is almost all subject to a high intensity of trawled gear fishing. The intensity and extension of fishing in the Adriatic Sea is unique at EU level, although the intensity increases on an East-to-West gradient. In the North East Atlantic the majority of the continental shelf is subject to trawling activities but with a high patchiness of intensity: coastal areas have greater intensity than off shore areas like the centre of the North Sea.

The classification of gear from the EU Fleet Register was assumed to be correct for selecting the towed gears. There might be however some misclassifications as there are some AIS fishing signals south of the Balearic Islands which are only compatible with gears Midwater Pair Trawl (PTM) and Midwater Otter Trawl (OTM), which are however not declared by Spain in 2014 according to the DCF data provided by Spain. Similarly in the Ligurian Sea there is an offshore group of AIS fishing signals that is not compatible with towed demersal gear and there is no official reporting of PTM and OTM in the area. Both these areas could be compatible with Purse seine or long line fishing. Based on these two cases there might some misclassification of Primary Gear in the Fleet Register that is worth investigating.

The level of detail of the layers allows identifying different types of fishing and can be used to investigate and characterise the relations between fishing grounds and fishing coastal communities for a coherent analysis of European fisheries from an environmental and socio-economics perspective.

A typical policy application of the map of fishing activity would be in the evaluation of the impacts of environmental conservation measures established on a geographical basis like in the case of fisheries management in Natura 2000 sites. Such measures include for example banning fishing activity using mobile bottom contacting gear (e.g. beam trawls and bottom otter trawls) to protect reef structures and achieving favourable conservation status under the Habitats Directive (for a recent example of policy application in the Baltic Sea and Kattegat see Commission Delegated Regulation (EU) 2015/1778). Since the protected sites are sometimes limited to few square km a detailed spatial representation of fishing intensity is needed in these cases to assess the environmental effectiveness of the planned measures. Knowing “who is fishing where” is essential to calculate indicators of spatial dependency of fishing fleets and coastal communities in the affected areas and these indicators can be easily converted into employment and GVA equivalents to assess and the socio–economic effects on the fishing sector.

In order to become a sound element of knowledge for researchers, public administrations and policy makers, the fishing intensity layer has to be associated to the level of AIS spatial coverage, which can be thought of as a reliability of information.

Fisheries management carefully keeps track of the distribution of fishing effort and its footprint. To get an estimate of fishing effort or of fishing footprint (Eigaard et al. 2015) it would be necessary to link the density of AIS with the size or the engine power of the fishing vessels, which is directly linked to the size of the gear towed. This additional step is beyond the scope of this paper but is a very important avenue for future research.

**Software**

We employ different software packages for the preparation of the map. AIS data were collated and cleaned using the Pandas data analysis library.

For the analysis part the statistical software R was used. R proved particularly useful and flexible but it required additional efforts to optimizing the code for computational efficiency. It was necessary to employ the latest software libraries available to R, above all “data.table” and parallelize the code.

The final results data were then converted into vector and raster data files and consequently mapped using Quantum GIS.

A georeferenced fishing intensity layers can be downloaded from the Blue Hub website, Mapping Fishing Activities (MFA) section.

**Acknowledgments**

The authors would like to thank MSSIS, courtesy of the Volpe Center of the U.S. Department of Transportation and the U.S. Navy, and MarineTraffic for providing the AIS data used in this study.

**Map Design**

The final map created is available to the public as a flat file and as an interactive layer hosted on the Blue Hub platform. The main layer in the map is the raster of fishing intensity, calculated as the number of messages classified as *'fishing'* per square kilometre. The mapping and design style used aim to simplify map reading by using a consistent set of fonts an easy-to-understand colour shading ramp and classification algorithms. The colour ramp adopted belongs to the famous *ColorBrewer* style in Quantum GIS, a colour scheme that has proven to be effective in conveying the intensity of phenomenon by using eye pleasing colours and giving the reader a less subjective interpretation of the intensity itself. We used a *'continuous'* algorithm to classify fishing intensity. The Map, contains also an inset with the coverage map and also a part showing the Gaussian Mixture Model speed profile used. The style of the map is very essential with a focus on grey yellow and blue colours. The choice of colours is aimed at people with colour vision impairments. The reference map is Natural Earth 1 with an overlay of a country boundaries vector layer in black with a light transparency to contrast the visual impact of the excellent Natural Earth dataset. No other boundaries have been added to hamper the readability of the map at the scale used, 1:7000000.